\documentclass{llncs}

\usepackage{times}
\usepackage[utf8]{inputenc}
\usepackage{booktabs}
\usepackage{tabularx}
\usepackage{graphicx}
\usepackage{ragged2e} 
\usepackage{amssymb}
\usepackage{amsmath}
 
\usepackage{listings}

\usepackage{listings, xcolor}

\definecolor{verylightgray}{rgb}{.97,.97,.97}

\lstdefinelanguage{Solidity}{
	keywords=[1]{anonymous, assembly, assert, balance, break, call, callcode, case, catch, class, constant, continue, contract, debugger, default, delegatecall, delete, do, else, emit, event, export, external, false, finally, for, function, gas, if, implements, import, in, indexed, instanceof, interface, internal, is, length, library, log0, log1, log2, log3, log4, memory, modifier, new, payable, pragma, private, protected, public, pure, push, require, return, returns, revert, selfdestruct, send, storage, struct, suicide, super, switch, then, this, throw, transfer, true, try, typeof, using, value, view, while, with, addmod, ecrecover, keccak256, mulmod, ripemd160, sha256, sha3}, 
	keywordstyle=[1]\color{blue}\bfseries,
	keywords=[2]{address, bool, byte, bytes, bytes1, bytes2, bytes3, bytes4, bytes5, bytes6, bytes7, bytes8, bytes9, bytes10, bytes11, bytes12, bytes13, bytes14, bytes15, bytes16, bytes17, bytes18, bytes19, bytes20, bytes21, bytes22, bytes23, bytes24, bytes25, bytes26, bytes27, bytes28, bytes29, bytes30, bytes31, bytes32, enum, int, int8, int16, int24, int32, int40, int48, int56, int64, int72, int80, int88, int96, int104, int112, int120, int128, int136, int144, int152, int160, int168, int176, int184, int192, int200, int208, int216, int224, int232, int240, int248, int256, mapping, string, uint, uint8, uint16, uint24, uint32, uint40, uint48, uint56, uint64, uint72, uint80, uint88, uint96, uint104, uint112, uint120, uint128, uint136, uint144, uint152, uint160, uint168, uint176, uint184, uint192, uint200, uint208, uint216, uint224, uint232, uint240, uint248, uint256, var, void, ether, finney, szabo, wei, days, hours, minutes, seconds, weeks, years},	
	keywordstyle=[2]\color{teal}\bfseries,
	keywords=[3]{block, blockhash, coinbase, difficulty, gaslimit, number, timestamp, msg, data, gas, sender, sig, value, now, tx, gasprice, origin},	
	keywordstyle=[3]\color{violet}\bfseries,
	identifierstyle=\color{black},
	sensitive=false,
	comment=[l]{//},
	morecomment=[s]{/*}{*/},
	commentstyle=\color{gray}\ttfamily,
	stringstyle=\color{red}\ttfamily,
	morestring=[b]',
	morestring=[b]"
}

\usepackage[backend=biber,style=numeric,citestyle=ieee,doi=false,isbn=false,arxiv=false]{biblatex}

\AtEveryBibitem{\clearfield{publisher}}
\AtEveryBibitem{\clearfield{editor}}
\AtEveryBibitem{\ifentrytype{article}{\clearfield{url}}{}}
\AtEveryBibitem{\ifentrytype{inproceedings}{\clearfield{url}}{}}
\AtEveryBibitem{\ifentrytype{book}{\clearfield{url}}{}}
\AtEveryBibitem{\ifentrytype{incollection}{\clearfield{url}}{}}
\DeclareSourcemap{
    \maps{
        \map{ 
            \step[fieldsource=url,
                  match=\regexp{\{\\\_\}|\{\_\}|\\\_},
                  replace=\regexp{\_}]
        }
        \map{ 
            \step[fieldsource=url,
                  match=\regexp{\{\$\\sim\$\}|\{\~\}|\$\\sim\$},
                  replace=\regexp{\~}]
        }
    }
}

\usepackage{todonotes}

\begin{document}

\title{Towards Safer Smart Contracts: A Survey of Languages and Verification Methods}
\author{Dominik Harz \and William Knottenbelt}
\institute{IC3RE \\ Imperial College London \\ \email{\{d.harz,wjk\}@imperial.ac.uk}}

\maketitle

\begin{abstract}
With a market capitalisation of over USD 205 billion in just under ten years, public distributed ledgers have experienced significant adoption.
Apart from novel consensus mechanisms, their success is also accountable to smart contracts. 
These programs allow distrusting parties to enter agreements that are executed autonomously.
However, implementation issues in smart contracts caused severe losses to the users of such contracts.
Significant efforts are taken to improve their security by introducing new programming languages and advance verification methods.
We provide a survey of those efforts in two parts.
First, we introduce several smart contract languages focussing on security features.
To that end, we present an overview concerning paradigm, type, instruction set, semantics, and metering.
Second, we examine verification tools and methods for smart contract and distributed ledgers. 
Accordingly, we introduce their verification approach, level of automation, coverage, and supported languages.
Last, we present future research directions including formal semantics, verified compilers, and automated verification.
\end{abstract}

\section{Introduction}
The idea of contracts between independent parties goes back to autonomous agents using a network of agents to solve tasks distributed and based on individual contracts as presented in the contract net protocol  \cite{Smith1980}.
The idea was further elaborated by Szabo using the term ``smart contract'', concentrating on minimising and ideally excluding the need for trust in either party,  \cite{Szabo1997}.
Significant work focussed on creating languages and frameworks for electronic contracts even before the inception of distributed ledgers, for example, \cite{Andersen2006,Kyas2008,Xu2004}.

Smart contracts based on distributed ledgers combine two unique properties: anyone can create such contracts for the whole world to interact with, while each line of code (LoC) might affect a significant amount of currency.
These contracts allow economic interactions between different parties without the need to trust one another.
Contracts are concerned with a range of use cases, includig financial services, notaries, games, wallets, or libraries \cite{Bartoletti2017}.
Further, smart contracts are the enabler of protocols build on top of distribued legers, for example, Lightning \cite{Poon2016}, Plasma \cite{Poon2017}, Polkadot \cite{Wood2017}, and TrueBit \cite{Teutsch2017}.
However, security incidents caused by software bugs have lead to severe losses as in the infamous The DAO incident \cite{Daian2016}, and Parity multi-sig vulnerabilities \cite{Breidenbach2017Parity,ParityTech2017}.

Substantial efforts are taken to prevent future incidents. 
High-level programming languages are introduced to encourage safe programming practices, for example  \cite{Hirai2018Bamboo,Ethereum2018Vyper,Schrans2018}.
Languages for distributed virtual machines that allow for easy verification are realised, for example \cite{Sergey2018,DynamicLedgerSolutions2017,Popejoy2017,Kasampalis2018}.
Tools for analysing source code by symbolic modelling and execution, for example \cite{Luu2016,Tsankov2017,Kalra2018,Albert2018} as well as formal semantics and verification, for example \cite{Bhargavan2016,Hildenbrandt2017,Hirai2017}, are developed.

\subsubsection{Contribution} The number of new languages, approaches for verification, and applicability of verification methods becomes quickly opaque. Due to the practical impact of these approaches to real-world smart contracts, we present a literature survey on current languages and verification efforts.
We contribute an overview of contract language security features including paradigm, instruction set, semantic, and metering.
Further, we describe different efforts to verify software including model- and proof-based methods. Our overview includes an analysis of the level of automation, coverage, and supported languages.

\subsubsection{Structure} The remainder of our article is structured as follows. Section \ref{background} introduces the background of contracts and languages to express them. We present an overview and a classification of languages in section \ref{languages}. Similarly, verification approaches are examined in section \ref{verification}. Results and future work is discussed in section \ref{discuss}. We conclude in section \ref{conclusion}.

\section{Safer smart contracts}
\label{background}



Contracts are a requirement resulting from an inherent lack of trust between parties.
Smart contracts execute agreements on a distributed ledger like Bitcoin or Ethereum \cite{Nakamoto2008,Buterin2013}.
The ledger's consensus protocol ensures correct execution of the smart contract, whereby a majority agrees on the accepted result.
Consensus protocols enable mutually \emph{distrusting parties} to create contracts and interact.
However, no single definition of a smart contract exists.
Information and definitions are scattered in academia and various communities (e.g.\ Bitcoin and Ethereum Improvement Proposals).
The capabilities of smart contracts range from restrictive (i.e.\ Bitcoin) to Turing-complete instruction sets (i.e.\ Ethereum). 
Whether or not smart contracts have legal implications or not is also widely discussed.
Moreover, there is a caveat with distributed ledgers and consensus protocols.
Results of contracts need to be fully deterministic so that each (honest) actor in the consensus protocol reaches a single common output with the same set of inputs.



\subsection{Technical overview}
Smart contracts need a low-level language that allows deterministic execution. 
A high-level language can make it easier for developers to create new contracts and reason about existing contracts.
In software development, having different sets of languages is a conventional process.
Similarly, we can distinguish three different levels of languages for smart contracts.

\subsubsection{High-level languages} They should provide a way to express the desired contracts. Multiple high-level languages can exist in parallel to be executed on the same ledger. Examples for high-level languages are Solidity \cite{Ethereum2018Solidity} and Liquidity \cite{OCamlProSAS2018}.

\subsubsection{Intermediary representations} IRs are between low-level and high-level languages. IRs can be used to write programs to reason about properties (like safety or liveness) or optimising code. Examples include Simplicity \cite{OConnor2017} and Scilla \cite{Sergey2018}.

\subsubsection{Low-level languages} These need to implement the contract in a deterministic way to be executed on a distributed virtual machine (VM). Examples include Bitcoin Script \cite{BitcoinWiki2018Script} and EVM bytecode \cite{Wood2014}.

\subsubsection{Distributed ledger} The ledger, implemented, e.g.\ as a blockchain or DAG, plays a vital role in the design of the language. 
The distributed ledger can be separated in transaction and scripts, consensus protocol, and network protocol. 
A more in-depth review is available from \cite{Bonneau2015}.
Bitcoin implements a UTXO model \cite{Nakamoto2008,Covaci2018}, where contracts are stored in the \texttt{scriptSig} and \texttt{scriptPubKey} of a transaction. 
Hence, contracts only cover a single transaction or need to be ``chained'' over multiple ones.
Account-based ledgers store contracts at a specific address.
Contracts have a local state, and their functions can be called multiple times.
State-changing functions are invoked by sending a transaction to a contract function exposed via its Application Binary Interface (ABI).
Contracts can typically access a global state to receive information stored in the ledger as block timestamps, block hashes, or transaction data. 



\subsection{Security properties}
Desired properties of smart contracts include safety and liveness \cite{Sergey2018}.
Safety refers to satisfying specific correctness properties during any state on a contract.
Liveness describes that certain events may eventually occur.
Analysis of vulnerabilities violating security properties with a focus on the EVM is presented in \cite{Luu2016} and \cite{Atzei2017}.
\citeauthor{Grishchenko2018} extend this work by introducing a formal definition of general security properties for smart contracts \cite{Grishchenko2018}:

\subsubsection{Call integrity}
A contracts state may depend on an external call. Specifically, a contract can execute code of an external contract within its functions, call another contract's function and wait for its returned value, or call another contract that changes the global state or re-enters the calling contract.
In these cases, the contract's control flow should not be influenced by an adversary contract.

\subsubsection{Atomicity}
Functions should be executed entirely, or the state should be reverted, except when specifically allowed during exception handling.
If a contract function has, for example, a \texttt{call} instruction not at the end of the function, the call might execute successfully, but the remainder of the function might terminate due to, e.g.\ out-of-gas error or a stack limit. Smart contracts are drained when a function sends currency and updates the balance after the send (which would not happen in case of an exception).

\subsubsection{Independence}
Transactions change the state of a contract. Miners and other parties can control or influence parameters in contracts or transactions. For example, a contract might use a timestamp-dependent function, which can in certain ranges be influenced by miners. Also, the ordering of transactions can be influenced by miners or others paying higher fees. Hence, contracts should ideally be independent of the global state or parameters that can be influenced externally.

\subsubsection{Runtime correctness}
The aforementioned properties apply to general contracts. However, each contract serves a distinct purpose. Properties particular to the expected behaviour of a contract need to be defined to ensure the runtime correctness of the contract.



%
%
%

\section{Contract languages}
\label{languages}

\begin{figure}
\label{fig:language}
\includegraphics[width=\textwidth]{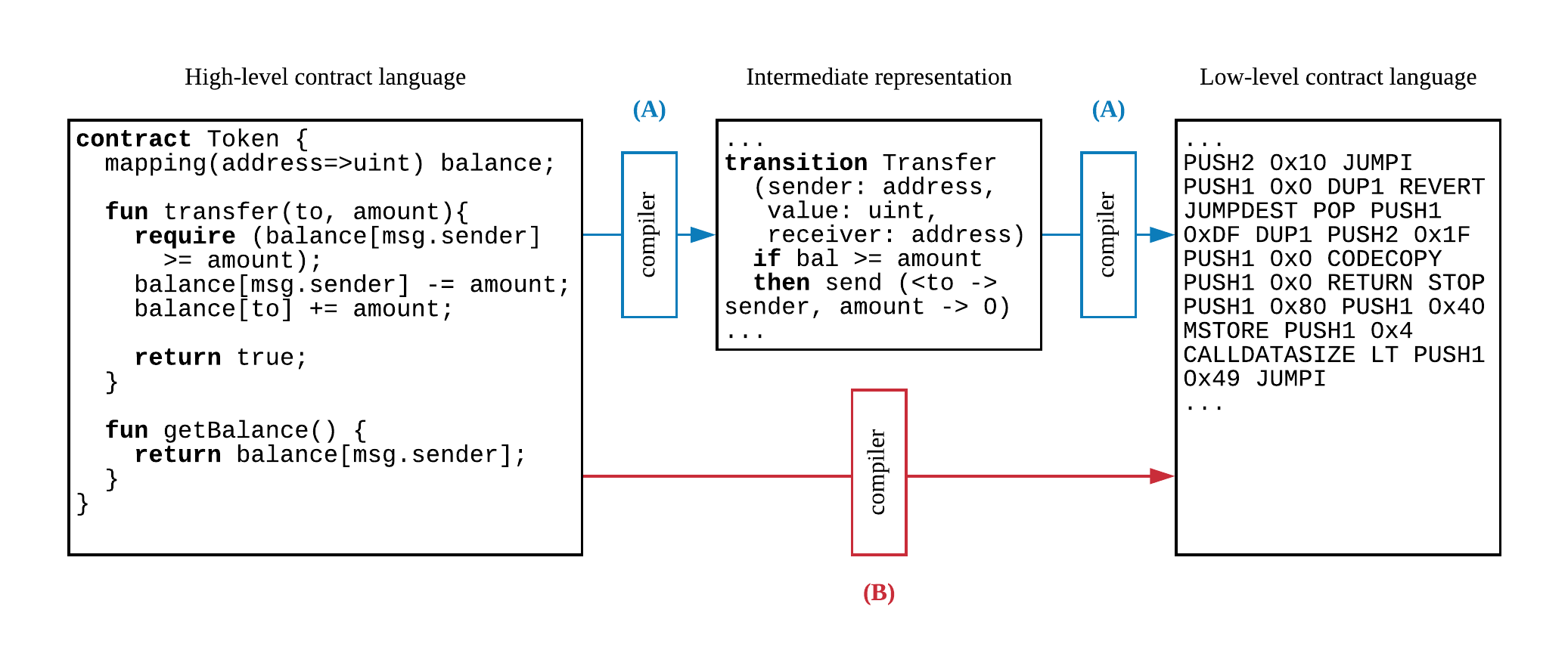}
\caption{Different levels of smart contract languages with syntax closely resembled to Solidity (high-level), Scilla (IR), and EVM bytecode (low-evel). (A) represents an optimised version of compiling high-level languages towards the bytecode that allows for example for verification of the IR contract and code optimisations, for example envisioned by \cite{Sergey2018,OCamlProSAS2018}. (B) represents the straightforward compilation from high-level language to bytecode representation as currently employed by Solidity and the EVM \cite{Ethereum2018Solidity,Wood2014}.}
\end{figure}

Figure \ref{fig:language} gives an overview of smart contract languages.
The approach in (A) is often used to allow program optimisation and verification (e.g. \cite{Lattner2004}). 
Among the projects that follow this approach is Ethereum with Yul \cite{EthereumFoundation2018IULIA}, and Tezos with Liquidity \cite{OCamlProSAS2018} and Michelson \cite{DynamicLedgerSolutions2017}.
Likewise, Scilla is an IR that is targeted by more general languages and compiles down to be executed on a distributed VM \cite{Sergey2018}.

Smart contracts are a comparably new discipline.
Hence, in the early stages, smart contracts were designed as represented by (B).
Bitcoin Script \cite{BitcoinWiki2018Script} requires programmers to write code directly in the low-level stack-based language.
Ethereum, on the other hand, offers multiple high-level languages like Solidity \cite{Ethereum2018Solidity} and Vyper \cite{Ethereum2018Vyper}.
These languages compile directly to EVM bytecode. 

\subsection{Overview}
The overview we present in table \ref{tab:high-level} is based on five different criteria as listed below. The table gives a general overview. We explain the security properties of the languages within the following subsections.
\begin{itemize}
\item \emph{Type}: We differentiate between high-level, IR, and low-level languages.
\item \emph{Paradigm}: This describes the main paradigm of the language. Note that most languages support multiple paradigms and this criterion is more of an indication of the prevalent paradigm.
\item \emph{Instructions}: The possible instruction set that a language supports can be restricted or Turing-complete.
\item \emph{Semantics}: Languages have a formal or informal semantic. Formal semantics define the exact behaviour of programs written in that language. Informal semantics mostly let the compiler define the exact behaviour.
\item \emph{Metering}: Smart contracts executed on a distributed ledger are re-executed by several nodes. As these computations are costly, metering is a way to charge and limit the execution of a program.
\end{itemize}
\begin{table}[h]
\centering
\caption{Overview of high-level languages for smart contracts.}
\label{tab:high-level}
\begin{tabularx}{\textwidth}{llXXXX}
\toprule
\textbf{Language} & \textbf{Paradigm} & \textbf{Instructions} & \textbf{Semantics} & \textbf{Metering} & \textbf{Ref.} \\ \toprule
\textit{Solidity} & contract-oriented & complete & informal & gas, limit & \cite{Ethereum2018Solidity} \\
\textit{Vyper} & procedural & restricted & informal & gas, limit & \cite{Ethereum2018Vyper} \\
\textit{Bamboo} & procedural & complete & semi\textsuperscript{\dag} & gas, limit & \cite{Hirai2018Bamboo} \\
\textit{Flint} & procedural & complete & informal & gas, limit & \cite{Schrans2018} \\
\textit{Pyramid Scheme} & functional & complete & informal & gas, limit & \cite{Burge2018} \\
\textit{Obsidian} & object-oriented & -- & informal & -- & \cite{Coblenz2017} \\
\textit{Rholang} & concurrent & complete & informal & phlogiston & \cite{Meredith2018} \\
\textit{Liquidity} & functional & restricted & semi\textsuperscript{\dag} & gas, limit & \cite{OCamlProSAS2018} \\
\textit{DAML} & functional & restricted & -- & -- & \cite{Shaul2018,Meier2018,Lippmeier2018,Huschenbett2018,Bernauer2018,Maric2018,Bleikertz2018,Lochbihler2018,Pilav2018} \\
\textit{Pact} & functional & restricted & semi\textsuperscript{\dag} & gas, limit & \cite{Popejoy2017} \\
 \midrule
\textit{Simplicity} & pure functional & restricted & formal & -- & \cite{OConnor2017} \\
\textit{Scilla} & functional & restricted & formal & gas, limit & \cite{Sergey2018} \\
\textit{Yul} & procedural & complete & informal & gas, limit & \cite{EthereumFoundation2018IULIA} \\
\textit{EthIR} & procedural & complete & informal & gas, limit & \cite{Albert2018} \\
\textit{IELE} & register-based & complete & formal & gas, limit & \cite{Kasampalis2018} \\ \midrule
\textit{Bitcoin Script} & stack-based & restricted & informal & script size & \cite{BitcoinWiki2018Script} \\
\textit{EVM} & stack-based & complete & informal\textsuperscript{\ddag} & gas, limit & \cite{Wood2014} \\
\textit{eWASM} & stack-based & complete & informal & gas, limit & \cite{EthereumFoundation2018ewasm} \\
\textit{Michelson} & stack-based & restricted & semi\textsuperscript{\dag} & gas, limit & \cite{DynamicLedgerSolutions2017} \\
\bottomrule
\end{tabularx}
\justify
\textsuperscript{\dag} These languages are actively developed. There are efforts to define a formal semantics. \\
\textsuperscript{\ddag} The EVM has been informally defined in \cite{Wood2014}. Formal semantics have been defined afterwards by \cite{Hirai2017,Hildenbrandt2017}.
\end{table}

\subsection{Languages}

\subsubsection{High-level}
Solidity is the most widely used smart contract language and created for Ethereum \cite{Ethereum2018Solidity}.
Bamboo is designed with formal verification in mind and makes state-transition explicit \cite{Hirai2018Bamboo}. 
Vyper restricts instructions (e.g.\ finite loops and no recursive calls) and prevents other features such as inheritance and overloading \cite{Ethereum2018Vyper}. 
Flint further introduces the definition of function access (by defining the address of the caller) and creates an asset type \cite{Schrans2018}. 
Pyramid Scheme is functional and imperative promoting separation of state-changing and static functions \cite{Burge2018}.

Obsidian models contracts as finite state machines (FSM) with explicit state transition functions \cite{Coblenz2017}.
Rholang focusses on concurrency and message-passing with statically typed communication channels \cite{Meredith2018}.
Liquidity has a restricted instruction set and enables formal verification \cite{OCamlProSAS2018}.
Rholang and Liquidity are intended for permissionless distributed ledgers.
DAML is functional and developed for financial applications, primarily on permissioned ledgers \cite{Shaul2018,Meier2018,Lippmeier2018,Huschenbett2018,Bernauer2018,Maric2018,Bleikertz2018,Lochbihler2018,Pilav2018}.
Similar, Pact is designed for the Kadena permissioned blockchain \cite{Popejoy2017}.
Both have a restricted instruction set and with the intention to promote formal verification.


\subsubsection{Intermediary}
Simplicity is a pure functional language that places itself as an intermediary representation between a higher level (functional) language and a low-level VM \cite{OConnor2017}. 
Scilla is functional with an automata-based design using explicit state transition functions and handling for communication patterns \cite{Sergey2018}. 
Yul (formerly IULIA and JULIA) is introduced as part of Solidity and its compiler \cite{EthereumFoundation2018IULIA}. 
EthIR is a decompilation target for EVM bytecode \cite{Albert2018}. 
IELE is derived from its formal semantics and used as an IR for smart contracts \cite{Kasampalis2018}. 
Scilla, Yul, EthIR, and IELE use an account-based blockchain model. Simplicity is built with a UTXO model in mind.

\subsubsection{Low-level}
Smart contracts are stored on a distributed ledger in the low-level language to be executed by the distributed VM.
Bitcoin scripts are a sequence of op-codes stored within transactions in the Bitcoin network \cite{BitcoinWiki2018Script}. 
The EVM stores programs in the data field of an address in the Ethereum network \cite{Wood2014}. 
eWASM is a proposed successor of the EVM based on a deterministic variant of Web Assembly (WASM) \cite{Wanderer2015,EthereumFoundation2018ewasm}.
Michelson is the low-level language of the Tezos blockchain \cite{DynamicLedgerSolutions2017}. It uses accounts as well but is designed to promote formal verification.

\subsubsection{General purpose languages}
Apart from using DSLs for programming smart contracts, projects like Hyperledger Fabric or Neo use general purpose programming languages.
This can have advantages as those languages are already known to potential developers, and verification tools might already exist.
For example, Hyperledger Fabric uses Docker containers with smart contracts (so-called ``chaincode'') written in Go, Java, or Node.js \cite{Cachin2016}. 

However, as these languages are originally not designed for smart contracts the global state of the ledger needs to imported through special functions that are typically not available in these languages.
Moreover, these languages often have support for infinite loops and recursion which are not desirable.
Particular types like assets or units also need to be defined appropriately. 

\subsection{Paradigm}

\subsubsection{Explicit state transitions}
Languages including Scilla, Rholang, Bamboo, and Obsidian as well as interfaces to Solidity \cite{Mavridou2018} represent contracts as finite state machines (FSM) or automata. This concept prevents reentrancy and allows to create explicit state transition function. A transaction that is sent to a contract with the intention to change the state either is successful or raises an exception. Moreover, this principle should prevent any calls to other contracts within a state transition function. A state might end or begin with a message (i.e.\ tail-call), but not have any external calls within the state that may change it unpredictably.

\begin{lstlisting}[caption={A separation of states represented in Bamboo, where each state represents a different contract at the same address.},label=lst:fsm,language=Solidity]
contract Voting() { ... }
contract VotingClosed() { ... }
contract Result() { ... }
\end{lstlisting}

\subsubsection{Functional programming}
Pyramid Scheme, Vyper, Simplicity, Scilla, and Bamboo as well as Pact and DAML for permissioned chains, use functional programming paradigms. Functions in these languages can be designed to be \emph{atomic} and execute entirely or revert. Also, pure functions can be used to indicate that the local or global state is not affected. Functions can call pure functions, but no other state-changing functions.

\subsubsection{Logic programming}
Logic-based languages are interesting as they closely resemble natural language contracts and have been explored in \cite{Idelberger2016}. Logic languages can be purposely non-deterministic. However, they transfer the burden of determinism to the low-level languages and the compiler.

\subsubsection{Stack-based}
All low-level languages are stack-based. Their low-level implementation makes a manual inspection of contracts cumbersome. Hence, automated tools can help to support such verification efforts. Moreover, decompilers are used to convert the stack into a higher level language.

\subsection{Instruction set}
\subsubsection{Restricted instructions}
Vyper, Liquidity, DAML, Pact, Simplicity, Scilla, Bitcoin Script, and Michelson restrict instructions, whereby Bitcoin Script is likely the most restrictive language (also considering that most op-codes are deactivated in Bitcoin). 
The idea is to prevent unwanted behaviour by restricting the instruction set to the necessary operations.
In practice, infinite loops and recursion would block any node in the network executing the smart contract. Hence, it can be directly restricted by the language.

\subsubsection{Tail-calls}
Another critical aspect of the instruction set is the possibility of calling other contracts. Executing code in other contracts can potentially introduce unexpected behaviour leading to unpredictable state changes. State changes can be made explicit by using an FSM or automata principle with \emph{tail-calls}. This principle is also suggested for Solidity \cite{ConsenSys2018Security}.
It offers a possibility to update the state of the contract and prevent potential adversaries to gain access to the contract control flow via reentrancy.

\begin{lstlisting}[caption={Tail calls implemented in Solidity.},label=lst:tail-call,language=Solidity]
function claim(uint id) public {
    require(msg.sender == owner[id]);
    funds[id] = 0;
    msg.sender.transfer(funds[id]);
}
\end{lstlisting}

\subsubsection{External function execution}
The EVM offers the \texttt{call}, \texttt{callcode}, and \texttt{delegatecall} instructions to interact with other contracts. The execution context changes based on the invoked call. When using special types within a function like \texttt{msg.sender} or \texttt{msg.value}, the instruction determines which transaction the types refer to. Moreover, \texttt{delegatecall} grants storage access from the calling contract to the called contract. An adversarial contract might thus change the calling contracts storage (i.e.\ local state) arbitrarily. Hence, these external calls should be restricted, or special communication types handle their correct execution as, e.g.\ in Scilla.


\subsubsection{Restrict overriding}
Overriding functions can lead to issues when reviewing code as it may not be clear which code is actually executed. Assume two functions as listed below.
\begin{lstlisting}[caption={Function overriding with different inputs.},label=lst:tail-call,language=Solidity]
function transfer(address to, uint amount) {}
function transfer(address from, address to, uint amount) {}
\end{lstlisting}
Both functions should theoretically implement the same behaviour. To prevent ambiguities languages like Vyper prevent function overriding.

\subsubsection{Overflow}
The EVM uses fixed-length integers of at most 256 bits.
Hence, developers typically use libraries (e.g. \texttt{SafeMath}) to prevent or detect overflows.
IELE, on the other hand, uses arbitrary-precision signed integers to prevent overflows altogether.


\subsubsection{Code re-use}
\citeauthor{Pontiveros2018} proposes to re-use EVM code to optimise the space usage of code with a new op-code \cite{Pontiveros2018}. This technique could be applied to reference proven secure code patterns.

\subsection{Semantics}

\subsubsection{Formal semantics}
Bamboo, Liquidity, Michelson, and Pact are planned to have a formal semantics once the languages are officially released.  Simplicity and Scilla have a formal semantics defined in Coq. IELE is defined in $\mathbb{K}$, and its implementation has been derived from the formal semantics.
Low-level languages have been informally defined. In later work, the EVM has been formally defined in $\mathbb{K}$ \cite{Hildenbrandt2017}, Lem \cite{Hirai2017}, and F* \cite{Grishchenko2018}.
As other languages have been informally defined, the correct interpretation of programs is left to the compiler. Creating a formal semantic for a higher-level language can enable the creation of verified compilers for said language and support verification efforts.

\subsubsection{Explicit types}
Flint introduces additional types for declaring assets. Smart contracts typically operate on fungible or non-fungible assets. Instead of using existing types (e.g.\ \texttt{uint}), particular asset types can provide additional safety measures like enforcing updating balances before using a \texttt{call} operation sending the asset.


\subsection{Metering}

\subsubsection{Script size limit}
Bitcoin Script uses a maximum script size of 10,000 bytes as a restriction to script complexity. Combined with its restricted instruction set, this enables liveness properties of the network. Merkelized Abstract Syntax Tree (MAST) is a proposal to allow larger scripts without increasing the script size limit \cite{Harding2017}.

\subsubsection{Gas}
The EVM, eWASM, IELE, and Michelson use gas and a gas limit to restrict instructions. Rholang's phlogiston essentially implements the gas concept as well. While this allows for liveness of the network as it restricts denial of service attacks, it also opens possible out-of-gas errors. For example, when a function terminates directly after completing a transfer without updating its state (like an account balance), the function can be invoked again without affecting the balance.
Hence, gas considerations are particularly important with external and untrusted calls.

\subsection{Additional security properties} 
\subsubsection{Contract upgrades}
Code is immutable in distributed ledgers. Hence, patterns using contract registries or \texttt{delegatecall} constructs have emerged to upgrade contracts. However, these patterns introduce problems. An alternative approach that sacrifises liveness for more safety is Hydra \cite{Breidenbach2018}.

\subsubsection{Randomness}
Since any local and global state needs to be deterministic and publicly verifiable in distributed ledgers, it is hard to obtain a source to generate pseudo-random numbers.
Miners can influence block-related parameters like timestamps and hashes. The RANDAO suffers from costly implementation. A proposed solution are verifiable delay functions \cite{Boneh2018}.

\subsubsection{Secrets}
As information needs to be verifiable in distributed ledgers, secrets are hard to implement. Commit-reveal schemes make use of hash functions where a user commits to a value by publishing a public hash of the value. Also, zero-knowledge proofs like zk-SNARKS are used to hide information while keeping them verifiable \cite{Sasson2014}.

\subsubsection{Best practices}
Apart from the features that a language supports, best practices can be applied to prevent unintended behaviours \cite{Wohrer2018,ConsenSys2018Security}.
Moreover, templates can be used to create smart contracts \cite{Clack2016}.

\section{Verification methods}
\label{verification}

\begin{figure}
\label{fig:verification}
\includegraphics[width=\textwidth]{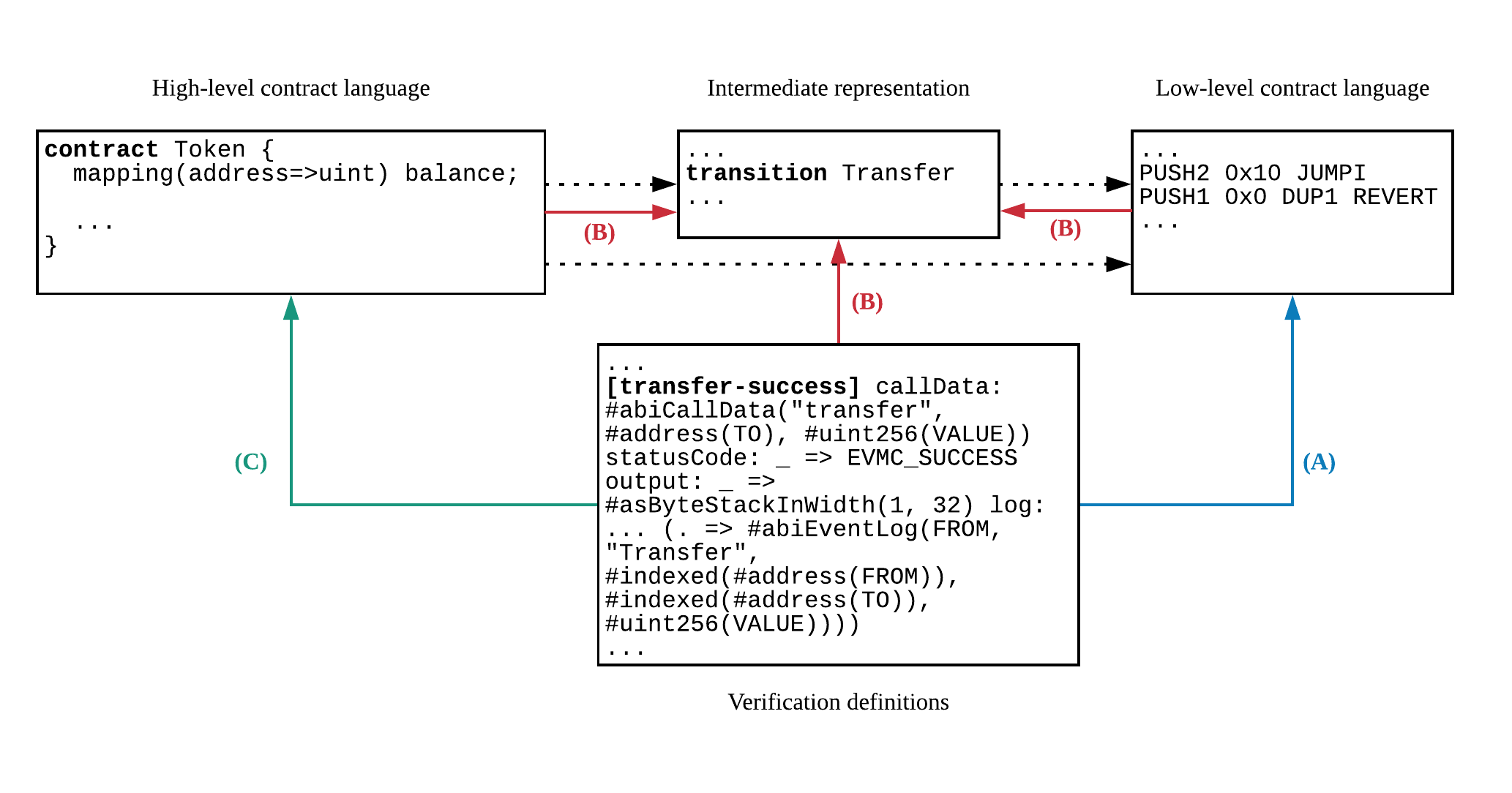}
\caption{Different approaches of verification tools regarding their source for deriving a model or a set of formulas representing the system. Methods listed in (A) directly verify properties from the model or set of formulas from the low-level code. (B) describes tools using an IR from a high-level language or from the low-level code to verify the system. Last, (C) describes methods for verifying properties directly from the high-level code.}
\end{figure}

Verification approaches are outlined in figure \ref{fig:verification}. 
(A) represents verification efforts that base their analysis on the low-level code that is or will be deployed on the distributed ledger. Those tools include $\mathbb{K}$/KEVM \cite{Hildenbrandt2017}, Lem (with their possible proofs in Coq or Isabelle/HOL) \cite{Hirai2017}, and F* \cite{Bhargavan2016,Grishchenko2018}.
Tools listed in (B) use the low-level code and decompile it into an IR to reason about properties in the contracts like Securify \cite{Tsankov2017}, Mythril \cite{Mueller2018}, Oyente \cite{Luu2016,Albert2018}, ECF \cite{Grossman2017}, and Maian \cite{Nikolic2018}. ZEUS is an exception as it uses a high-level language to compile an IR and is not verifying properties based on the low-level code \cite{Kalra2018}.
(C) describes tools that reason directly on the high-level code. Solidity can be annotated with Why3 statements to reason about the correctness of the contract \cite{Reitwiessner2015Why3}. Oyente can be used as well, but will compile code into an IR. Those methods can help to find bugs in contracts, but since they do not operate directly on the low-level code, they rely on verified compilers to infer properties like safety or liveness.

\subsection{Overview}

Table \ref{tab:model} presents an overview of the tools we have considered for this survey.
Verification efforts can be characterised by five criteria \cite[173]{Huth2004}. From these, we adopt three and fixate the other two. 
The application domain concerns smart contracts on a deterministic distributed ledger.
On these ledgers smart contracts are \emph{immutable}, hence, verification before or during, but latest, before deployment, is desirable. Otherwise, they can only be used to find existing bugs and requires significant efforts to update the contract.
Additionally, we included the language that is covered by the specific tool as well as the availability of source code.
This leaves us with the following criteria.

\begin{itemize}
\item \emph{Approach}: In proof-based verification, the system is represented by a set of formulas, while in model-based verification the system is a model. Properties are represented as formulas. The goal is to either proof (proof-based) or to compute whether a model (model-based) satisfies these properties. Proof-based methods typically derive a formal definition of the distributed VM and then try to verify properties of a smart contract. Model-based methods build a model directly from the smart contract and verify the properties with an implicit model of the VM.
\item \emph{Automation}: Fully automated approaches have a set of properties and automatically build a model for the system based on an input (like the source code). Partial automation typically requires defining properties or using a proof-assistant (e.g. Coq or Isabelle/HOL) to define and check proofs.
\item \emph{Coverage}: Property-based verification is concerned with selected parts of the system, while full covers the system as a whole.
\item \emph{Languages}: Lists the languages that are currently, as of September 2018, supported by the methods. Some of the general tools like Lem, F*, $\mathbb{K}$, and Coq can be used for any languages.
\item \emph{Source}: Indicates whether the tools and the verification work is available as open source. This criterion is interesting to verify results, experiment with the available tools, and potentially expand them.
\end{itemize}

\begin{table}
\centering
\caption{Overview of model and proof-based verification tools for smart contracts.}

\label{tab:model}
\begin{tabularx}{\textwidth}{XXXXlXX}
\toprule
\textbf{Tool} & \textbf{Approach} & \textbf{Automation} & \textbf{Coverage} & \textbf{Languages} & \textbf{Source} & \textbf{Ref.} \\ \toprule
\emph{Securify} & model & full & full & Solidity, EVM & open & \cite{Tsankov2017} \\
\emph{Mythril} & model & full & property & EVM & open & \cite{Mueller2018} \\
\emph{Oyente} & model & full & property & Solidity, EVM & open & \cite{Luu2016,Albert2018} \\
\emph{ZEUS} & model & full & property & Solidity, Go, Java & closed\textsuperscript{\dag} & \cite{Kalra2018} \\
\emph{ECF} & model & full & property & EVM & open & \cite{Grossman2017} \\
\emph{Maian} & model & full & property & EVM & open & \cite{Nikolic2018} \\ \midrule
\emph{$\mathbb{K}$} & proof & partial & full & EVM, IELE & open & \cite{Hildenbrandt2017,Park2018} \\
\emph{Lem} & proof & partial & full & EVM & open & \cite{Hirai2017,Amani2018} \\
\emph{Coq} & proof & partial & partial\textsuperscript{\ddag} & Scilla, Michelson & open & \cite{Sergey2018,DynamicLedgerSolutions2017} \\
\emph{F*} & proof & partial & partial\textsuperscript{\ddag} & EVM & open & \cite{Bhargavan2016,Grishchenko2018} \\
\bottomrule
\end{tabularx}
\justify
\textsuperscript{\dag} We were not able to find open source code. \\
\textsuperscript{\ddag} Theoretically these tools have a full coverage. However, implementation is as of September 2018 not completed.
\end{table}

\subsection{Tools}
Model-based methods arise from the need to check contracts for known vulnerabilities. After the DAO and Parity vulnerabilities were known, tools were developed to find similar patterns in other contracts. The proof-based methods arise from the need to proof contracts secure. This requires a formal semantics of the VM and the low-level language. Overall, these methods are used to prevent future vulnerabilities but depend on an exact definition of properties and rigorous formal semantics.

\subsubsection{Model-based}
Securify is a domain-specific model checker for smart contracts \cite{Tsankov2017}. It compiles EVM bytecode to semantics facts and then uses a dialect of Datalog to define compliance and violation properties to verify the semantic facts. It classifies behaviours of a contract in compliance (matched by compliance properties), violations (matched by violation properties), and warnings (matched by neither). 
Mythril is a symbolic execution of EVM bytecode \cite{Mueller2018}. EVM bytecode is disassembled into a Mythril object, and propositional logic is used to reason about the state space represented as a graph. 
Oyente \cite{Luu2016} and its proposed extension EthIR \cite{Albert2018} build a model from EVM bytecode to verify pre-defined properties. Properties include transaction ordering dependencies, timestamp dependencies, mishandled exceptions, and reentrancy.
ZEUS uses Solidity or Java and Go (as Hyperledger Fabric contracts) as its basis for evaluation \cite{Kalra2018}. It compiles these contracts into an abstract language. Next, properties defined in XACML are used to reason about the contract. The properties together with the abstract language contract get translated to LLVM bitcode for symbolic execution and verification of the properties.
Effectively Callback Free (ECF) objects are a property that is analysed for Ethereum smart contracts\cite{Grossman2017}. A callback method opens up the possibility to change the state of an object (contract) from an external object (contract), which makes reasoning difficult. 
Maian works by symbolic execution of a model of EVM bytecode contracts to find trace vulnerabilities \cite{Nikolic2018}. These vulnerabilities include contracts that leak Ether to unintended parties, can be killed by arbitrary users or lock Ether that cannot be received. 

\subsubsection{Proof-based}
$\mathbb{K}$ is a general purpose framework for defining programming languages \cite{Rosu2007}. It is used to build a $\mathbb{K}$ representation of the EVM, called KEVM \cite{Hildenbrandt2017}. 
A contract can be formally verified using the compiled bytecode, the $\mathbb{K}$ contract, and the KEVM virtual machine. 
Lem is used to defining language semantics and can be used to derive implementations in OCaml and enable proof-based verification using Coq, HOL4, or Isabelle/HOL \cite{Mulligan2014}. The EVM has been defined in Lem and subsequently contracts verified using the semantic definition \cite{Hirai2017}. This work is extended by \cite{Amani2018}. 
Coq is an interactive theorem prover that can be used for any language theoretically. In practice, Scilla is defined in Coq, and there are ongoing efforts to verify Scilla contracts \cite{Sergey2018}. Further, Coq is intended to be used together with the Michelson language \cite{DynamicLedgerSolutions2017}.

\citeauthor{Bhargavan2016} propose to convert Solidity and EVM bytecode to F* \cite{Bhargavan2016}. This can then be used to verify properties in the contract and obtain a secure implementation. However, the work does not present a full implementation.
Further, a complete small-step semantics of the EVM semantics is presented in \cite{Grishchenko2018}. Based on this semantics the authors have implemented in large parts the EVM in F*. F* has then been compiled to OCaml code to verify the EVM implementation against the official Ethereum test suite.

\subsection{Automation} 
\subsubsection{Full automation}
Model-based tools are automated. They usually use an SMT solver (e.g. Z3) to explore the fulfilment of violation of properties. Automation offers a significant advantage as the pre-defined properties in the tool can easily be verified on other contracts. Moreover, Securify, Oyente, and Mythril are available as a web-service. This allows developers to check their contracts without the need to install dependencies for model checking locally.

\subsubsection{Partial automation}
Proof-based tools are partially automated. The Lem, $\mathbb{K}$, F* semantics are focussing on creating the distributed VM that executes the smart contracts. Automation can be reached by defining properties contracts should fulfil. This is partly done by the ERC20 efforts in $\mathbb{K}$ and the ``Deed'' contract in Isabelle/HOL. However, it is desirable to define the functionality of a contract and then verify its safety, correctness, and liveness rather than finding selected vulnerabilities. The verification of the properties is then done using an SMT solver (in $\mathbb{K}$) or using an interactive theorem prover (Isabelle/HOL, Coq, F*).

\subsection{Coverage} 
\subsubsection{Partial, property-based verification}
Most model-based tools verify selected properties. In \cite{Atzei2017} and \cite{Luu2016}, the authors offer a classification of possible vulnerabilities. These vulnerabilities build the basis for the properties to check as the tools try to identify violations or conformance of those patterns to flag a contract as vulnerable.
Model-based tools rely on detecting these properties by static analysis. 
Oyente and Mythril are shown to miss vulnerable patterns (false negative) and flag safe contracts as vulnerable (false positive) \cite{Tsankov2017}.

\subsubsection{Full verification}
Securify gives a warning if none of its conformance or violation patterns matches.
The desired coverage of proof-based methods is the full contract. 
General smart contract security semantics are formally defined in \cite{Grishchenko2018}. They build a basis for the F* small-step semantics and can be adopted to other general proof-based techniques as well.
However, the security semantics presented are not complete.
Additional, contract specific, properties need to be defined to ensure correctness of the program.
By giving a formal specification of the contract functionalities, a contract can be deemed correct. This approach is beneficial for common standards (e.g. ERC20 or ERC721). A short-coming of proof-based verification is that a verified contract might contain bugs due to incomplete or inaccuracies in the specification or VM semantics \cite{Hirai2016}.

\subsection{Languages} 
\subsubsection{EVM}
The majority of the tools use the EVM bytecode to derive a model of the contract. Moreover, most models do not implement all EVM opcodes. Hence, vulnerabilities might remain undetected as not all contracts can be fully verified.
Major work efforts are taken in building a formal semantics of the EVM ($\mathbb{K}$, Lem, F*). 
The F* implementation is partially complete. 

\subsubsection{IR and low-level languages}
While the EVM semantic is defined after its implementation, IELE, Scilla, and Michelson are designed with formal verification in mind. Hence, their semantics are currently developed and in the future formal verification should be comparably easy. Further, their formal semantics approach helps to build verified compilers.

\subsubsection{High-level languages}
ZEUS is an exception as it builds the model based on higher-level languages such as Solidity or Java and Go. 

\subsection{Source} 
\subsubsection{Documentation}
All tools have a description in a paper or technical report that gives details about their internals. They offer extensions by creating new properties for verification.
The $\mathbb{K}$ framework has extensive documentation and examples available, followed by the work on Lem and Isabelle/HOL. The Coq and F* methods have been introduced this year, and documentation is yet sparse.

\subsubsection{Source code} Except for Securify and ZEUS, tools can be cloned locally, and additional properties can be added.

\section{Discussion}
\label{discuss}

\subsubsection{Language design}
High-level languages for smart contracts are designed and improved to promote safe smart contracts.
This is achieved by making state changes explicit by using an FSM/automata approach. 
They typically restrict the instruction set by only allowing finite loops and recursion.
Moreover, the code should be as explicit as possible by preventing function overloading, creating explicit types for assets and units, and promoting pure functions.
Intermediary languages are developed with formal verification and optimisations in mind.
This is an attempt to bring best practices from software engineering and language theory to distributed ledgers.
Low-level languages are built to allow formal verification and at the same time give a run-time optimised execution on a distributed ledger.
Combining those practices and applying them in the development cycle helps create secure smart contracts.

\subsubsection{Verification}
Verification efforts include categorising and defining security properties for smart contracts, developing model-based tools to verify that contracts are not vulnerable to known bugs, and formal semantics with the intention to prove compliance of a contract implementation to an abstract specification. Proof-based verification requires more effort than model checking for existing vulnerabilities. Hence, verification is sensible for high-value and critical contracts. In fact, the Casper contract is currently being formally verified.

\subsubsection{Formal semantics and verified compilers}
A main focus is developing IRs with formal semantics and creating formal semantics for existing low-level languages. We argue that in the future, more projects need to adopt formal semantics on all language levels to promote verification efforts and prevent ambiguities in compiler implementations. Further, this allows to create verified compilers making it easier to argue about contracts in a high-level language \cite{Hirai2017}. 
Next, current formal semantics for the EVM combine the VM and the ledger in a single definition. Future work is to separate the semantics for the ledger and the execution environment. 

\subsubsection{Complete security definitions}
An initial proposal of formal security definitions is made in \cite{Grishchenko2018}. Those definitions are taken from the perspective of Solidity and the EVM. Hence, more general security definitions for various execution environments are required. Also, it could be interesting to separate the execution environment from the ledger. Moreover, proposed standards like the ERC20 can be formally defined as part of the proposal process. This would allow to verify implementations against a formal specification. Possibly, automated formal verification methods can be built on this.





\section{Conclusion}
\label{conclusion}
An overview of smart contracts and verification methods is presented.
Languages are developed and improved to allow easier verification by defining formal semantics. Also, secure patterns like explicit state transitions and restricted instructions are applied. Verification efforts concentrate on finding known vulnerabilities and formally defining smart contract logic to verify implementations.

We note three areas of future work. Formal semantics are being adopted for low-level languages, and selected IRs have been designed from a formal semantics. However, high-level languages with full formal semantics are just being developed.
Formal semantics on all levels of languages is a requirement to develop verified compilers. Verified compilers and formal semantics can then be used to build automated proof-based verification methods.


\subsubsection{Acknowledgement}
We like to thank Ben Livshits for his helpful comments on our work. This research is funded by the Outlier Ventures research grant for the Imperial Centre for Cryptocurrency Research and Engineering.

\printbibliography

\end{document}